\documentclass[authoryear]{elsarticle}
\usepackage[latin1]{inputenc}
\usepackage{amsmath}
\usepackage{amsfonts}
\usepackage{amssymb}
\usepackage{natbib}

\begin{document}
\title{The Necessity of the Second Postulate in Special Relativity}

\author{Alon Drory}
\ead{adrory@gmail.com}
\address{Afeka College of Engineering, 218 Bney-Efraim Street, Tel-Aviv, 69107 , Israel}

\begin{abstract}

Many authors noted that the principle of relativity together with space-time homogeneity and isotropy restrict the form of the coordinate transformations from one inertial frame to another to being Lorentz-like. The equations contain a free parameter, $k$ (equal to $c^{-2}$ in special relativity), which value is claimed to be merely an empirical matter, so that special relativity does not need the postulate of constancy of the speed of light. I analyze this claim and argue that the distinction between the cases $k = 0$ and $k \neq 0$ is on the level of a postulate and that until we assume one or the other, we have an incomplete structure that leaves many fundamental questions undecided, including basic prerequisites of experimentation. I examine an analogous case in which isotropy is the postulate dropped and use it to illustrate the problem. Finally I analyze two attempts by Sfarti, and Behera and Mukhopadhyay to derive the constancy of the speed of light from the principle of relativity. I show that these attempts make hidden assumptions that are equivalent to the second postulate.
\end{abstract}

\begin{keyword}
Special Relativity  \sep Lorentz Transformations \sep constancy of the speed of light \sep Second Postulate \sep Electromagnetism
\end{keyword}

\maketitle

\section{Introduction}
\label{intro}
In 1905, Albert Einstein published what became eventually known as the theory of special relativity (SR). He based it on two principles, or postulates \citep{einstein1}:

\begin{quote}
1. The laws governing the changes of the state of any physical system do not depend on which one of two coordinate systems in uniform translational motion relative to each other these changes are referred to.

2. Each ray of light moves in the coordinate system ``at rest'' with the definite velocity V independent of whether this ray of light is emitted by a body at rest or in motion. \citep[p. 895, English translation from The Collected Papers, Vol. 2, 1989, p.143]{einstein1}
\end{quote}

Einstein implicitly assumed also that space and time are homogeneous and isotropic, a fact quickly pointed out and stressed by several writers. These symmetries seem so natural, however, that SR is still said to be based on Einstein's two postulates: the principle of relativity (postulate 1) and the principle of the constancy of the speed of light (postulate 2).

Almost immediately, the necessity of the second postulate was questioned \citep{ignatowski1, ignatowski2, frank}. Recognized as fundamental and deep, the principle of relativity seemed appropriately general for such a fundamental theory. Contrariwise, the second postulate struck many investigators as too particular and contingent to merit its elevated position \citep{weinstock,mitvalsky,levy-leblond,srivas,schwartz1,sen,pal}. 

In 1965, for example, Robert Weinstock thought that:
\begin{quote}
...the status of the theory of relativity would be rendered somewhat more secure if it were to be based on an experiment less accessible to interpretative controversy than that of Michelson and Morley. \citep[p. 641]{weinstock}
\end{quote}

His own suggestion was to replace the second postulate by another experimental fact, viz:

\begin{quote}
The crucial experimental result on which the theory of relativity is based in this paper is the nonconstancy of the mass of a body as a function of its speed relative to the inertial frame in which the mass is measured. \citep[p. 641]{weinstock}
\end{quote}

Most other writers on the subject, on the other hand, feel that a second postulate - \textit{any} kind of second postulate - is unnecessary. Thus, A.M. Srivastava (1981) begins his derivation of the Lorentz transformations by stating that: 

\begin{quote}
It has been pointed out by many authors that the postulate of the constancy of the speed of light is not necessary for arriving at the space-time transformations in special relativity \citep[p. 504]{srivas}
\end{quote}

Similarly, L\'{e}vy-Leblond (1976 )declares:

\begin{quote}
...I intend to criticize the overemphasized role of the speed of light in the foundations of the theory of special relativity, and to propose an approach to these foundations that dispenses with the hypothesis of the invariance of c.\citep[p. 271]{levy-leblond} 
\end{quote} 

David Mermin (1984) expressed most clearly the reason behind these concerns :

\begin{quote}
Relativity .... is not a branch of electromagnetism and the subject can be developed without any reference whatever to light.\citep[p. 119]{mermin}
\end{quote}

I believe that here lies the source of the discontent with the second postulate. Special relativity is one of the deepest theories of physics, a revolutionary reworking of our understanding of space and time - the fundamental stage on which the physical universe is played out. Surely, such a fundamental theory must proceed from similarly fundamental and deep postulates. Yet while the principle of relativity possesses the required generality, the second postulate appears to be too incidental and specific to play such a fundamental role. As Mermin points out, the second postulate seems to reduce special relativity to being a mere branch of electromagnetism rather than a fundamental theory from which electromagnetism can actually be derived by applying the Lorentz transformations to electrostatics \citep[for an approach to electromagnetism along these lines, see for example, ][]{schwartz}. 

It seems to me that this feeling is the reason why the necessity for the second postulate has been repeatedly challenged over the years, and so much effort dedicated to trying to prove that SR follows from space-time symmetries and the principle of relativity alone. In particular, various derivations of the Lorentz transformations along those lines obtain as a \textit{result} the existence of an invariant (and maximal) velocity. 

I shall name ``generalized Lorentz transformations'' the coordinate transformations obtained from the spacetime symmetries and the principle of relativity alone, in order to distinguish them from the transformations used in SR that make direct reference to the speed of light. The generalized transformations contain an undetermined universal parameter, here denoted $k$, which is equal, in standard SR, to $\dfrac{1}{c^2}$. In these views, the content of the second postulate is no more than a report of the experimental value of $k$, or equivalently, of the maximal invariant velocity, which happens - for no particular reason, one might think in the context of these derivations - to be the velocity of light. By this I mean that light itself appears to play no particular role in these presentations; the identification of the invariant velocity with that of light is viewed as being a claim of electromagnetism, but one not inherent to SR itself.

These various derivations are undoubtedly of great interest. That the mathematical form of the possible coordinate transformations is as restricted as it is by the relativity principle and the space-time symmetries is hardly obvious and remains surprising to the physicists and philosophers who are unaware of this result. Other insights also originated from similar reflections, such as the role of isotropy in SR in Feigenbaum's own attempt to derive SR from relativity alone \citep{feigen}. The abstract of Feigenbaum's paper reads in part ``No reference to light is ever required: The theories of relativity are logically independent of any properties of light'', a view very close to the one espoused by Mermin.

Yet despite these interesting insights, I believe that these derivations fail in their avowed aim. My position is not new. Such was already the opinion of Pauli, quite early on \citep{pauli}. Having quickly reviewed the derivation of the generalized Lorentz transformations, he wrote:

\begin{quote}
Nothing can naturally be said about the sign, magnitude and physical meaning of [k]. From the group-theoretical assumption, it is only possible to derive the general form of the transformation formulae, but not their physical content.  \cite[p.11]{pauli}
\end{quote}

Unfortunately Pauli offered no elaboration of this claim, which led many subsequent investigators to think they had managed to overcome this criticism. I believe this is not the case, and that indeed the content of the second postulate (in one form or another) cannot be reduced to the mere experimental determination of some parameter in a theory derivable from the principle of relativity alone. Instead, as Pauli noted, the constancy of the speed of light is necessary in order to make \textit{physical} sense of the theory (as opposed to mere formal structure), particularly of what might count in it as ``experimental determination''. There are several aspects to this claim, and the present paper is intended to be the first in a series exploring the various roles fulfilled by the second postulate \citep{drory2}.

Section \ref{sec:rules} sets the ground rules for my analysis. Section \ref{sec:trans} presents a derivation of the generalized Lorentz-transformations, which is in my opinion simple and direct enough to justify its inclusion here. In section \ref{sec:galileo}, I argue the the non-vanishing of $k$ must be taken as an extra-assumption, i.e., another postulate. Sections \ref{sec:behera} and \ref{sec:sfarti} critique two attempts to derive the actual value of $k$ (including its non-vanishing) without assuming it, i.e., attempts to prove the second postulate. The final section presents a summary of the arguments. An appendix contains the derivation of anisotropic transformations that are used as an analogy, in the analysis in section \ref{sec:galileo}.

\section{Ground rules}
\label{sec:rules}

What is the nature of the claims made by supporters of the one postulate vision? In order to evaluate properly the value of the arguments and their problems, we must first agree on some basic rules.

First, the claims that are put forwards have no relation to the historical processes leading up to the discovery of the theory of relativity. That the second postulate was historically crucial to Einstein's thinking is of no consequence here. It is the logical structure of the theory, assessable in retrospect, which is under discussion. Thus the proper aim seems to be the following:

Assume only the following three postulates:

A.	Homogeneity of space and time: The laws of physics are invariant under a translation of the origin of coordinates of space and time. 

B.	Isotropy of space: The laws of physics are invariant under rotations of the axes in which they are described.

C.	The principle of relativity: The laws of physics are invariant when referred to different frames of reference, when these are in uniform motion with respect to each other.

It is traditional to consider the principle of relativity to imply the law of inertia, although formally it should be treated as a separate axiom. The rationale for this inclusion is apparently that one takes for granted something along the lines of a principle of sufficient cause - specifically, that a body at rest that is sufficiently isolated from other bodies (and hence from external influences) remains at rest. The property ''being at rest'' is not invariant to different observers, however, and if one accepts the rest principle and the principle of relativity, one should uphold the more general formulation that an isolated body in uniform motion will also maintain that state. 

For the present purpose, however, it matters little whether one takes the principle of inertia to be a supplementary axiom or a consequence of the principle of relativity (together with some principle of sufficient cause). The principle of inertia is assumed in all derivations of the generalized Lorentz transformations as well as in the derivations of the standard transformations, so that its inclusion neither adds nor detracts anything from the one-postulate argument.

The aim of one-postulate supporter is now to derive, from these assumptions alone, the special relativistic kinematics up to a numerical constant that is determinable in principle \textit{without invoking further physical knowledge}.

The italicized part above is essential to the evaluation of the enterprise. By claiming that the theory of relativity, at least its kinematics, can be derived from just the assumptions mentioned above, we bar ourselves from using additional theoretical or empirical data. Were it not the case, one could conceivably claim that the principle of relativity itself need not be counted among the theory's assumptions, since it can be obtained by appeal to established tradition, mechanics and /or countless experiments, going back to the time of Galileo and Descartes. 

Of course, historically, that principle needed reiterating in its general form because Maxwell's electrodynamics seemed to conflict with it. But under the rules of the game we are trying to play, electrodynamics must also be ignored. Were this not the case, the second postulate could be said to follow from Maxwell's theory just as well, which is of course the way Einstein approached his derivation. 

Thus, what is at stake here is whether we can derive in a meaningful way special relativistic kinematics, from just the three assumptions A-C.

\section{Derivation of the generalized Lorentz transformations}
\label{sec:trans}

Assume a frame of reference S, with space and time coordinates $(t,x,y,z)$, and a second frame S' moving relative to S with a velocity $v$ directed along the x axis. We seek the relation between the coordinates $(t,x,y,z)$ of an event in the S frame and the coordinates $(t', x', y', z')$ of the same event when observed in S'. Let us write the general coordinate transformation as:

\begin{eqnarray}
\label{general}
x' = F(x,t) \nonumber \\
t' = G(x,t)
\end{eqnarray}

The claim that these are the general transformations is not trivial. For starters, prior to the theory of relativity, the notion that time was not invariant was unthinkable, and the second expression would therefore not even have been considered. Furthermore, it is possible that the transformations ought to include other parameters that seem irrelevant today but might be proven important by some future, more general, theory. The decision to write down the transformations in this form is therefore itself a product of SR. While this has no impact regarding the logical structure of the theory, it is nevertheless a significant fact.

Now consider two events, separated by a distance $dx$ and a delay $dt$. Let the first event's coordinates be $(t_0, x_0, y_0, z_0)$, and the second's $(t_0 + dt, x_0 + dt, y_0, z_0)$. The separation between these two events as observed from S' is now

\begin{eqnarray}
\label{interval}
dx' = \dfrac{\partial F(x_0,t_0)}{\partial x_0}dx + \dfrac{\partial F(x_0,t_0)}{\partial t_0}dt \nonumber \\
dt' = \dfrac{\partial G(x_0,t_0)}{\partial x_0}dx + \dfrac{\partial G(x_0,t_0)}{\partial t_0}dt
\end{eqnarray}

We now invoke the homogeneity of space and time, i.e., invariance under translation of the origin of space and time. This property implies that the space-time interval observed in S' cannot depend on the values of $x_0$ and $t_0$. Since $dx$ and $dt$ are independent of each other, this requires that all the partial derivatives in Eq.(\ref{interval}) are constant. Hence, we find immediately that the transformations are linear.

Next consider the origin of S', the point $x' = 0$. We define the relative velocity of the frame S' by requiring that in S, the coordinates of this point obey the relation $x = vt$ (here I assume that the origins coincide at the moment $t = t' = 0$). Hence, we can write the transformations as:
\begin{eqnarray}
\label{stage1}
x' = \gamma\left( x - vt \right) \nonumber \\
t' = \alpha \left(t - \epsilon x \right),
\end{eqnarray}
where $\alpha, \epsilon, \gamma$ are as yet undetermined constants, which may depend on the parameter $v$. 

To determine the form of this dependence, we make use of spatial isotropy. The direction chosen to be the positive $x$ axis is arbitrary, and the transformation ought to be unchanged if we reverse the direction of this axis, so that $x \longrightarrow -x$ and $x' \longrightarrow -x'$. We do not reverse the sense of the motion of S', but since this motion was in the positive $x$ direction before, after the reversal it will be in the negative direction of the new $x$ axis. Hence, to express the fact that the physical situation is unchanged, we must also transform $v \longrightarrow -v$. After this reversal, the transformation between the new spatial coordinates becomes
\begin{equation}
-x' = \gamma(-v)\left[-x - (-v)t\right]  \Longrightarrow  x' = \gamma(-v)\left(x - vt\right)
\end{equation}

Comparing this result with Eq.(\ref{stage1}) shows immediately that

\begin{equation}
\gamma(v) = \gamma(-v)
\end{equation}

Similarly, the transformation for the time coordinate becomes, under reversal of the positive direction of the $x$ axis:
\begin{equation}
t' = \alpha(-v)\left[ t - \epsilon(-v)\cdot(-x)\right]
\end{equation}

Comparison with Eq.(\ref{stage1}) now shows that

\begin{eqnarray}
\alpha(-v) = \alpha(v)\nonumber \\
\epsilon(-v) = - \epsilon(v)
\end{eqnarray}

These equations can of course be used not only to express an inversion of the $x$ axis, but also to transform to a frame of reference moving with a velocity $-v$ rather than $v$, since the behavior of $\alpha, \epsilon, \gamma$ under a reversal of the sign of $v$ is independent of whether we also reverse the $x$ axis or not (because of the space-time symmetries we assume). Using this fact, we can now combine two such transformations; one from $S_1$ to a frame $S_2$ moving at a velocity $v$, and from $S_2$ to a frame $S_3$ moving at a velocity $-v$. This frame ought then to coincide with the original frame $S_1$, so that $x_3 = x_1$ and $t_3 = t_1$, where
\begin{eqnarray}
\label{inverse}
x_3 = \gamma(-v)\left(x_2 + vt_2 \right) \nonumber \\
t_3 = \alpha(-v)\left[t_2 - \epsilon(-v)x_2 \right]
\end{eqnarray}

If we substitute the expressions for $x_2$ and $t_2$ from Eq.(\ref{stage1}) into these equations, we obtain

\begin{eqnarray}
\label{combine}
x_1 = x_3 = \gamma(-v)\left[\gamma(v) - v\alpha(v)\epsilon(v)\right]x_1 + v\gamma(-v)\left[\alpha(v) - \gamma(v)\right]t_1 \nonumber \\
t_1 = t_3 = \alpha(-v)\left[\alpha(v) + v\gamma(v)\epsilon(-v)\right]t_1 - \alpha(-v)\left[\epsilon(v)\alpha(v) + \epsilon(-v)\gamma(v)\right]x_1
\end{eqnarray}

Since $x_1$ and $t_1$ are independent, the left hand side and the right hand side of the equations will be equal only if

\begin{eqnarray}
\label{relatio}
\alpha(v) = \gamma(v) \nonumber \\
\gamma^2(v)\left[1 - v\epsilon(v)\right] = 1
\end{eqnarray}

Define a new parameter $\phi (v) $ by

\begin{equation}
\label{defphi}
\phi(v) = v\epsilon(v)
\end{equation}

Then, Eq.(\ref{relatio}) simplifies to

\begin{equation}
\alpha(v) = \gamma(v) = \dfrac{1}{\sqrt{1- \phi(v)}}
\end{equation}

This leaves only the parameter $\phi(v)$ to be determined. To do this, we imagine three different reference frames, S , S' and S''. S' moves at velocity $v$ with respect to S, and S'' moves at velocity $dv$ with respect to S', where $dv$ is infinitesimal. The combined transformation yields now:
\begin{equation}
\label{group1}
x'' = \gamma(dv)\left(x' - t'dv\right) = \gamma(dv)\gamma(v)\left[\left(1 + \epsilon(v)dv\right)x - \left(v+ dv \right)t\right]
\end{equation}

Here we have used Eq.(\ref{stage1}) to express $x', t'$ through $x, t$. Now, $\gamma(0) = 1$, since $v = 0$ means that there is no change in the space and time coordinates. Moreover, since $\gamma(v) = \gamma(-v)$, we must also have that 
\begin{equation}
\dfrac{d \gamma(v)}{d v}\bigg\vert_{v = 0} = -\dfrac{d \gamma(-v)}{d v}\bigg\vert_{v = 0}  \Longrightarrow  \; \; \dfrac{d \gamma(v)}{d v} \bigg\vert_{v = 0} = 0
\end{equation}

Thus, we see that $\gamma(dv) = 1 + O\left( dv^2 \right)$ and to first order, we can take $\gamma(dv) \approx 1$. To first order in $dv$, Eq.(\ref{group1}) now becomes
\begin{equation}
\label{group2}
x'' = \gamma(v)\left(1 + \epsilon(v)dv\right)\left[x - \dfrac{v + dv}{1 + \epsilon(v)dv}t\right] + O(dv^2)
\end{equation}

Now, this equation must express the direct transformation of coordinates from frame S to frame S''. Let us denote by $w$ the velocity of the frame S'' as seen from the frame S. Then, the direct coordinate transformation must have the form $x'' = \gamma(w)\left( x - wt \right)$. Comparison with Eq.(\ref{group2}) shows that we must have
\begin{eqnarray}
\label{group3}
w = \dfrac{v + dv}{1 + \epsilon(v)dv} = v + \left[1 - v\epsilon(v)\right]dv + O(dv^2) = v + \left[1 - \phi(v)\right] dv + O(dv^2) \nonumber \\
\gamma(w) =  \gamma(v)\left[1 + \epsilon(v)dv\right] + O(dv^2)
\end{eqnarray}

To the same order in $dv$ we also have that 

\begin{equation}
\label{group4}
\gamma(w) = \gamma(v) + \dfrac{d \gamma(v)}{d v}\left(w - v \right) + O(dv^2) =  \gamma(v) + \dfrac{d \gamma(v)}{d v}\left[1 - \phi(v)\right]dv + O(dv^2)
\end{equation}

Equating Eq.(\ref{group3}) and Eq.(\ref{group4}) yields the following differential equation:

\begin{equation}
\dfrac{d \gamma(v)}{d v}\left[1 - \phi(v)\right] = - \gamma(v)\epsilon(v)
\end{equation} 

Expressing all the parameters in terms of $\phi$ finally yields, after some simple algebra,
\begin{equation}
\dfrac{d \phi (v)}{dv} = 2\dfrac{\phi(v)}{v}
\end{equation}

This is immediately integrable and yields $\phi(v) = kv^2$ where $k$ is a universal constant, independent of $v$. Hence, also, $\epsilon(v) = kv$. We can now write the final form of the coordinate transformations:
\begin{eqnarray}
\label{final}
x' = \dfrac{\left( x - vt \right)}{\sqrt{1 - kv^2}} \nonumber \\
t' = \dfrac{\left( t - kvx \right)}{\sqrt{1 - kv^2}}
\end{eqnarray}

Note that the two left hand sides of Eq.(\ref{group3}) for the combined velocity $w$ are exact and do not assume that $dv$ is infinitesimal. Denoting the velocity of S'' with respect to S' by $u$ (to make clear it is a finite quantity) we find immediately the rule for the addition of velocities:
\begin{equation}
\label{veladd}
w = \dfrac{v + u}{1 + \epsilon(u)} = \dfrac{v + u}{1 + kuv}
\end{equation}

The remaining question is the value of the parameter $k$, which has units of $(velocity)^{-2}$. 

With one additional bit of (almost trivial) empirical information, Eq.(\ref{veladd}) shows that $k$ must be non-negative. If we had $k < 0$, we could combine two positive velocities $u, v$ and obtain a negative velocity $w$, provided $uv >  \dfrac{1}{\vert k \vert}$. Now if $k$ is negative, the denominator in Eqs.(\ref{final}) never vanishes and there is no limit velocity (although $\vert k\vert ^{-1/2}$ is an invariant velocity, but not the maximal possible one). Therefore the condition $uv >  \dfrac{1}{\vert k \vert}$ is clearly possible and obtaining a negative $w$ should also be possible. This means the following. Imagine an observer in a train rolling a ball on the floor towards the head of the train, say to the right. To an observer on the ground, watching the train itself move to the right, the ball might appear to move leftwards if $k < 0$. In the end however, the ball must reach the front of the train, which itself moves right, so that this seems to be a contradiction. Whether the contradiction is strictly logical or ``merely'' a violation of the most trivial empirical observations is rather unimportant. Clearly, $k$ cannot be negative under standard kinematics. 

If $k$ is non-negative, the rule of velocity addition, Eq.(\ref{veladd}), shows that $k^{-1/2}$ is an invariant velocity, usually denoted $c$. Srivastava (1981) expresses what appears to be the general opinion when he writes:

\begin{quote}
As we know, the experiments show that $c$ has a finite value which is equal to the value of the speed of light. \citep[p. 505]{srivas}
\end{quote}

In this view, the equality appears to be purely contingent, and the value of $k$ might conceivably have been anything else, as far as the present derivation is concerned. Electromagnetism obviously provides grounds to relate $k$ to the speed of light, but the logic of the present derivation is meant to exclude any appeal to that theory. Under such a conception, light holds no fundamental position in the theory - not only because other signals move at the same speed, so that the value of $k$ could be derived from, say, gravitational waves detection - but rather because no property of light enters essentially in the derivation of the equations. David Mermin (1984) goes farther and thinks that:

\begin{quote}
It is not necessary for there to be phenomena propagating at the invariant speed [$k^{-1/2}$] to reveal the value of $k$. One only requires sufficiently accurate measurements of the speed of any uniformly moving object from two different frames of reference. \citep[p. 123] {mermin}
\end{quote}

The notions of experiments and measurements require a great deal of care, however. I shall argue elsewhere that in trying to give meaning to the experimental determination of $k$, the second postulate (or something like it) turns out to be essential \citep{drory2}. In the present work, on the other hand, I shall concentrate on the question of the non-vanishing of $k$.

\section{Between Galileo and Lorentz}
\label{sec:galileo}

The generalized Lorentz transformations, Eqs.(\ref{final}), look very similar to the standard Lorentz transformations, and it is tempting indeed to say that we have thus derived special relativity (up to the numerical determination of $k$) without assuming the constancy of the speed of light. 

But have we indeed obtained a theory similar to special relativity? I wish to concentrate on the question whether $k$ is non-zero or not. The vanishing or non-vanishing of $k$ goes far beyond the mere report of a numerical measurement. The case $k = 0$ represents a fundamentally different theory from the case $k \neq 0$. The former is Newtonian mechanics, the second a relativity-like theory (with SR itself being specifically the theory that corresponds to $k = 1/c^2$). Yet nothing in the postulates A-C of the generalized-Lorentz transformations determines which of the two cases applies. I wish to argue that until we do in fact decide whether $k$ vanishes or not, we have not yet derived a complete physical theory because several fundamental questions remain undecided. 

\subsection{An Anisotropic Theory}

The argument is delicate because SR is so familiar that it is hard to forget what we know about it and look at it afresh as a ``new'' theory in which all we know are the assumptions A-C and their consequences. Because of this, I wish to present an analogy. Let us imagine for rhetorical purposes that we are dealing with a different approach, namely an attempt to derive special relativity from the following postulates:

E.	Homogeneity of space and time.

F.	The principle of relativity.

G.	The principle of constancy of the speed of light.

These assumptions differ from those of standard special relativity by not including the isotropy of space (in fact, postulate E - the homogeneity of space-time - turns out to be redundant, but for the rhetorical point I am trying to make, it does not matter). These turn out to be insufficient to derive the Lorentz transformations themselves, but they do imply that the transformations between two frames of reference must be of the form:

\begin{eqnarray}
\label{finalx}
x' = \left( \dfrac{1 - \beta}{1 + \beta} \right)^{\eta} \dfrac{x - \beta c t}{\sqrt{1 - \beta^2}}  \\
\label{finaly}
y' = \left( \dfrac{1 - \beta}{1 + \beta} \right)^{\eta} y \\
\label{finalz}
z' = \left( \dfrac{1 - \beta}{1 + \beta} \right)^{\eta} z \\
\label{finalt}
ct' = \left( \dfrac{1 - \beta}{1 + \beta} \right)^{\eta} \dfrac{ct - \beta x}{\sqrt{1 - \beta^2}}
\end{eqnarray}
where $\beta = \dfrac{v}{c}$, as usual.

The derivation of this result appears in the appendix. These equations differ from the Lorentz transformations by the inclusion of a new parameter, $\eta$, which represents a coefficient of spatial anisotropy. I show in the appendix that $\eta$ depends in general on the orientation of the x-axis, and that there is always a special direction in space along which $\eta = 0$. Letting $\theta$ be the angle between this special direction and the $x$-axis of the system, it turns out that $\eta$ depends on this angle and verifies the relation $\eta(\theta + \pi) = -\eta(\theta)$ (see the appendix). 

I shall refer to Eqs.(\ref{finalx})-(\ref{finalt}) as the anisotropic transformations for ease of reference. Now just like the generalized Lorentz equations, these contain an undefined numerical parameter ($k$ in the generalized Lorentz transformations, $\eta$ here). In both cases, the theory says nothing about the value of these parameters and they are left to empirical determination. The anisotropic transformations are unfamiliar, however, which makes them a good testing ground for the importance of the vanishing/non-vanishing distinction.

The Lorentz transformations are contained in the anisotropic ones. We are therefore merely an experiment away from special relativity - the empirical determination of $\eta$. A value of $\eta = 0$ will leave us with special relativity. And yet, clearly we would not say on this ground that special relativity is derivable without the postulate of spatial isotropy. The distinction between a zero and non-zero value of $\eta$ is not a purely numerical matter, but something fundamental to the theory.

Consider first the possible discovery that $\eta$ is non-vanishing, therefore. The importance of this result does not depend on whether in some direction $\eta = 0.2$ or 20. The numerical values are important to specific applications, but they make little difference to the world-view that would emerge from such a result. On the other hand, the qualitative discovery that our world is anisotropic rather than isotropic has far-reaching consequences. 

First of all, the philosophical implications are considerable, of course. That our world might lack some fundamental symmetry is an important alteration to our conception of the universe. 

Second, there would be serious theoretical implications. Many derivations use isotropy as an ingredient; these would have to be revised. Noether's theorem, for example, establishes the conservation of angular momentum from isotropy. Discovering that $\eta \neq 0$ would either imply that the conservation law is broken in general (though approximately valid since obviously any non-vanishing value of $\eta$ in our world must be small) or imply a different expression for the quantity angular momentum itself. Similarly, the symmetry groups that limit the form of future theories (more specifically, e.g., their Langrangians) would differ from the ones used nowadays.

Thirdly, a non-vanishing value of $\eta$ would have implications for the practice of experimental physics, in particular as regards the set-up of experiments and recording of their results. If our world is anisotropic, one must record the precise direction in which every experimental component is facing and repeating an experiment would demand that the experimental set-up be oriented in the same way as the original. At present, however, such precautions are not taken as the prevailing opinion is that these details are unimportant because our space is isotropic (anisotropic effects that depend on the rotation of earth, such as the Coriolis force, must of course be taken into account even now). Furthermore, past results would need to be re-evaluated to make sure that their conclusions are not tainted by anisotropy effects. Our analysis of astronomical observations, in particular, would have to be reviewed. 

In view of this, the case $\eta = 0$ must represent far more than merely a different empirical value for some constant. The entirety of the effects mentioned above would simply disappear. The question is not one of approximations. If $\eta$ were non-vanishing but small enough, many experiments could dispense with a careful recording of the orientation of the apparatus, but if $\eta$ is known to vanish, the question would not even be raised. Thus, the very questions we ask, the procedures we follow, the definitions we use - in short, the very structure of the theory - would all be different. 

But even after we accept that the vanishing or non-vanishing of $\eta$ is a fundamental question of the theory, could we not still leave it to experimental determination?

Of course the value of $\eta$ should be established empirically. But the vanishing/non-vanishing distinction is fundamentally different to just measuring, say, the value of Planck's constant to one more decimal point. That distinction must be part of the postulates of the theory.

Suppose, for example, that our experiments indicated a possible vanishing of $\eta$. No experiment can distinguish a precise value of zero from a sufficiently small finite value, so no direct measurement of $\eta$ can ever yield the result that it is exactly zero. All we can expect is to set upper bounds on its value and deduce that the experiments are compatible with a vanishing result. 

In that case, then, special relativity proper would be the theory obtained by \textit{adding} the postulate of spatial isotropy to the principle of relativity and the principle of the constancy of the speed of light. Thus, the claim that $\eta = 0$ would have to be raised to the level of a separate principle, even if only because experiments can only suggest upper bounds. In fact, there are other reasons why $\eta = 0$ deserves recognition as a separate postulate, which will become clearer as we proceed. This does not mean that it becomes impervious to empirical verification, quite the contrary. What it means is that special relativity as a theoretical structure is distinguished by the assumption that $\eta$ vanishes. 

Clearly, then, we cannot claim that postulates E-G allow us to derive special relativity. \textit{That} theory requires a supplementary postulate because we could never establish empirically the vanishing of $\eta$. But are we not justified then in claiming that postulates E-G are the basis of a different theory, one in which space is anisotropic? I shall claim that the answer is negative as well. 

Suppose that we found that $\eta$ is non-vanishing. Clearly, the theory represented by the anisotropic transformations is no longer special relativity, and this fact must be acknowledged somehow. The new theory would receive a different appellation, no doubt, and would be rightly considered incompatible with classical special relativity. This distinction hinges merely on the fact that $\eta$ is non-vanishing, not on any specific value it might have. If two sets of experiments pointed to two different values of $\eta$, both non-vanishing, this would not represent two competing theories, but merely an empirical indistinctness, to be resolved by more precise measurements. The theory would be considered the same in both cases, since its qualitative predictions would remain unaffected.

What makes this theory deserve the distinction of a separate name would be precisely that its predictions qualitatively differ from those of special relativity on a fundamental level. As noted above, it entails different preconditions of experiments. It requires different data to assess the meaning of an experiment, such as noting the orientation of every component of the apparatus. It has different implications for other theories, such as cosmology. It represents a different fundamental symmetry group for future physical theories. All these differences depend on the non-vanishing of $\eta$, not on its specific value. Thus, the fact that $\eta \neq 0$ becomes sufficient to classify the theory as a new one. 

But then why are we not allowed to claim that postulates E-G do in fact form the basis of this different theory? The answer is that just as the vanishing of $\eta$ is a fundamental aspect of special relativity, so is the non-vanishing of $\eta$ a fundamental characteristic of the new theory. And the fact is, that nothing in postulates E-G implies that $\eta \neq 0$.

This is a crucial point, not mere hair-splitting. We have just seen that the case $\eta = 0$ and $\eta \neq 0$ correspond to two \textit{different} theories, by the  accepted custom of what deserves to be called a physical theory. Yet both option are contained in postulates E-G, with nothing to allow us to separate them. Of course experiments will determine which of the two holds in our world (at least they will if $\eta \neq 0$; a vanishing value could only be suggested as a tentative conclusion). But every information we have about the world ultimately connects with experiments in some way. Just because a piece of knowledge comes to us empirically does not mean that it should not be taken as a postulate when discussing the logical structure of a theory. Indeed, what is the second postulate of special relativity itself but just such an empirical result raised to the level of a postulate? The exact same thing can be said of the laws of thermodynamics, and of just about any physical theory.

To distinguish the anisotropic theory from special relativity requires adding to its postulates the empirical fact that $\eta$ does not vanish. It would not be a postulate in the sense of the previous two, i.e., not a general principle. It would be ``merely" an empirical datum, but that datum would have to be considered part of the fundamental structure of the theory because that is what distinguishes it from an incompatible competitor, namely, special relativity. 

One may counter that $\eta \neq 0$ ought to be the ``default'' assumption when interpreting the anisotropic transformations, Eqs.(\ref{finalx}). This would certainly be the practice while the theory is developed. After all, had Eqs.(\ref{finalx}) been actually presented as a new physical theory and put to the test, its supporters would obviously consider it worthy of interest only if $\eta \neq 0$. Any other assumption would simply make it uninteresting and therefore unworthy of further investigation.

But explaining what drives a specific individual (or community) to spend time and energy on a proposal is not the same as setting out a theory in a logical fashion. In practice, the assumption that the postulates E-G imply anisotropy would be made \textit{implicitly} because that would be the only practical justification for investigating it. But implicit or not, it still represents an \textit{additional} assumption, and when, in retrospect, we investigate the logical structure of that theory, we are bound to make explicit all the ``natural'' additions that are taken for granted in the heat of discovery. The same holds, after all, of space-time homogeneity and isotropy themselves in the development of special relativity. They were not set as explicit postulates in Einstein's original paper, but were always added to the list when the logical structure of the theory was presented in orderly fashion. ``Natural'' and ``obvious'' as these postulates may be, no one doubts that they \textit{must} be added to the list of assumptions on which special relativity rests.

The fact is that the anisotropic transformations do not represent \textit{a} theory, they represent \textit{two} separate ones: special relativity (which corresponds to $\eta = 0$) and a new theory ($\eta \neq 0$), which we do not hold at present, and contains a different physics. As physical theories, as logically structured claims about the world, these two differ in their assumptions, although both verify postulates E-G. The correct conclusion is not that postulates E-G generate a different theory but that they are insufficient to determine \textit{one specific} physical theory. They are only part of the postulates needed for that. The structure to which they belong can only deserve the name of physical theory when an additional assumption is added, whether explicitly or implicitly, whether determined purely empirically or driven by abstract principles, and that assumption must be the distinction between vanishing or non-vanishing values of $\eta$. 

\subsection{Special Relativity vs Newtonian Mechanics}

This brings us back, finally, to special relativity itself and to the generalized Lorentz transformations. By now, the analogy should be clear. As a physical theory, special relativity contradicts Newtonian mechanics. It implies that the Newtonian world-view is inadequate. The Galilean transformations are of course a formal limit of the Lorentz transformations ($ c \rightarrow \infty)$, but the physical contents of the theories are incompatible. We may neglect any particular numerical effect under specific circumstances, but the theory of special relativity, qua physical theory, leaves no opening for the possible correctness of the Galilean transformations. 

If we were to discover tomorrow that photons have mass and that therefore light does not in fact propagate with the velocity $\dfrac{1}{k^{1/2}}$, many details would be different, but the world-view of relativity would remain untouched. The limiting speed would no longer be called the speed of light, but this would not change the fundamentals of the theory. Not so were we to bizarrely discover that $k = 0$ (though of course, here as well no experiment can decide that $k$ strictly vanishes, only that it is smaller than some bound). Then we would likely declare that special relativity is dead, and that \textit{another} theory is correct - Newtonian mechanics, which is fundamentally incompatible with it.

This means that just like the hypothetical anisotropic theory mentioned above, the foundations of special relativity contain something more than just the formal structure of the coordinate transformations (in this case, the generalized Lorentz transformations). As a physical theory, it is distinguished from its competitors by the claim that in our world $k$ is non-vanishing. 

Of course, we know that this is the case. In \textit{our} world, $k$ is indeed not zero, and whatever the future status of special relativity, a return to Newtonian mechanics is impossible. That the principle of relativity and the space-time symmetries should be valid but that $k$ should vanish is simply impossible in our world as we understand it. But this is hindsight, and it has become so natural to us that we are likely to miss the fact that the non-vanishing of $k$ is a peculiarity of special relativity that sets it against Newtonian physics and apart from it, a peculiarity that is not implied in any way by postulates A-C.

As with the case of $\eta$, the vanishing of $k$ would certainly deserve to be raised to the level of a postulate because that is the defining principle of a different theory. Newtonian physics is just as distinguished by special relativity via the additional principle that $k = 0$, as special relativity is distinguished from the anisotropic theory by the principle that $\eta = 0$ (which is equally inaccessible to experiments and can only be pronounced compatible with empirical data but never definitely confirmed by it). Conversely, just like an anisotropic theory has to be distinguished from special relativity by the added claim that \textit{in fact} $\eta \neq 0$, so must special relativity be distinguished from Newtonian mechanics as an independent theory by the fact that $k \neq 0$. This piece of information is part of the very foundations of the theory and is an essential part of its identity qua physical theory.

The generalized Lorentz transformations make no such distinction, however, and in this sense, they cannot be said to represent special relativity anymore than they represent Newtonian mechanics, though they are compatible with both theories (unlike the standard Lorentz transformations). Precisely because these transformations contain the possibility of Newtonian physics as well, by themselves they leave the world-view of the theory undefined. Thus, the structure they embody can only be considered as a fully determined physical theory once a decision has been made on the question of the vanishing of $k$. $k = 0$ defines one physical theory; $k \neq 0$ defines a different type of physical theory, one that shares a fundamental ontology with special relativity, though it may differ from it in details if $k$ turns out not to be equal to $\dfrac{1}{c^2}$. As with the case of the anisotropic theory, this distinction must be made explicit, and as such it becomes one of the postulates of the theory. 

Obviously, the second postulate sets special relativity against Newtonian mechanics in a way that the derivation of the generalized Lorentz transformations simply does not. What this derivation shows, and it is an important result, is that special relativity is in some sense the \textit{maximal} theory allowed by the principle of relativity and the space-time symmetries. There are no more free parameters that can be juggled around except $k$. Thus, this particular set of postulates allows just two types of physical theories: Newtonian mechanics and special relativity. It cannot distinguish between them but it does forbid anything more general than special relativity and this is by itself a very interesting insight; but this should not confuse us into thinking that special relativity has been thereby selected as \textit{the} theory implied by these postulates. It is \textit{a} theory implied by them, and it is only distinguished from its competitor by the additional postulate that $k \neq 0$. Although more limited than Einstein's second postulate (which also connects the value of $k$ to a specific physical phenomenon), it is nevertheless an additional postulate required by the theory before it can be said to yield definite information about the world. 

To be explicit, consider the following set of questions that may be asked of the ``theory'' obtained from postulates A-C alone, without the constancy of the speed of light. According to these postulates, are lengths identical for all inertial observers? The correct answer is neither yes nor no, but rather ``either'', depending on whether $k$ vanishes or not. Is the rate of clocks identical for all inertial observers? Is there a finite limit velocity that cannot be reached or crossed? Is simultaneity agreed upon by all inertial observers? To these questions, all fundamental to the philosophical world-view of the theory, to its implications regarding other theories and their allowed symmetry groups, and finally to physical practice (should we note the velocity of the object whose length we just measured or not?) - to all these questions, postulates A-C cannot offer a definite answer. They only suggest that ``it depends''. Once again, this is not to be disparaged. The notion that there is even a possibility that such effects arise is an important realization, but it remains merely a possibility until we come down clearly on one side or another of the issue of the vanishing of $k$.

In short, just like the anisotropic transformations, the generalized Lorentz transformations do not represent a single theory, but rather two contradictory ones: Newtonian mechanics ($k = 0$) and a class of special relativity-like theories ($k \neq 0$), distinguished in their quantitative predictions by the exact value of $k$ but all sharing a common world-view and identical philosophical, theoretical and practical implications for physics. As with the case of the anisotropic transformations, the distinction between these two cases is not merely a numerical matter. These are two different classes of theories, not one theory with an undetermined parameter, and by both current and historical practices and attitudes, we consider them as separate views of the world, yet postulates A-C by themselves are insufficient to separate them. Only with the addition of another postulate concerning the possible values of $k$ can we claim to have established a true physical theory.

\section{The derivation of Behera and Mukhopadhyay}
\label{sec:behera}

In view of the last section, it is interesting to note that there were two attempts to derive special relativity proper (as opposed to the generalized Lorentz transformations) without assuming the constancy of the speed of light. I analyze one of them in this section, the other in the next.

In a paper from 2003, Behera claimed to be able to derive the second postulate from the first \citep{behe}. That paper contained an unwarranted assumption that the shape of a light wavefront was spherical in all frames of reference. This assumption was dropped in the later work of Behera and Mukhopadhyay \citep{behera}. Because the remainder of the argument is identical in both papers, I will consider only the later work. My remarks apply equally to the 2003 paper, however, since the exact same mistake mars both derivations. 

In their derivation of the Lorentz transformations, Behera and Mukhopadhyay assume linearity, and write the transformations in the form

\begin{eqnarray}
\label{behera1}
x' = k(x - vt) \; ; \; y' = y \; ; \; z' = z \nonumber \\
t' = lx + mt  
\end{eqnarray}
where $k, l, m$ are constants to be determined \citep{behera}. To this end, Behera and Mukhopadhyay imagine that a light signal is emitted from a point source at rest at the origin of the S frame. The isotropy of the system ensures that the light propagates with an identical speed, c, in all direction. The wavefront of the emitted signal is spherical, therefore, and an arbitrary space-time point on the wavefront, $E = (ct, x, y, z)$, verifies the relation

\begin{equation}
\label{waveS}
x^2 + y^2 + z^2 = c^2t^2
\end{equation}

In the S' frame \citep[denoted F' in][]{behera}, the speed of light is $c'$. Behera and Mukhopadhyay then correctly note that

\begin{quote}
[W]e are yet uncertain about the shape of the same wavefront in [S'] because we know not yet whether c' is uniform in all directions of [S'] or not. \citep[p. 178]{behera}
\end{quote}

They note that $t'$ is the same for all points on the wavefront, however. Consequently, the coordinates of the arbitrary point $E$ on the wavefront are $(ct', x' , y', z')$ and they verify the relation

\begin{equation}
\label{waveSp}
x'^2 + y'^2 + z'^2 = c'^2t'^2
\end{equation}

Behera and Mukhopadhyay then substitute Eq.(\ref{behera1}) into Eq.(\ref{waveSp}) and obtain

\begin{equation}
\label{behera2}
\left( k^2 - l^2c'^2 \right)x^2 + y^2 + z^2 -2\left(lmc'^2 + k^2v\right)xt = \left(m^2c'^2 - k^2v^2\right)t^2
\end{equation}

So far all is well. Now comes the crucial step. Behera and Mukhopadhyay claim that in order for Eq.(\ref{behera2}) to be consistent with the description of the wavefront event E in the S frame, Eq.(\ref{waveS}), we must have that 

\begin{eqnarray}
\label{behera3}
k^2 - l^2c'^2 = 1 \nonumber \\
lmc'^2 + k^2v = 0 \nonumber \\
m^2c'^2 - k^2v^2 = c^2 
\end{eqnarray}
From these relations, they easily derive the form of the transformations. An application of the reciprocity condition (essentially, that the transformations themselves should be covariant) then yields that $c = c'$. 

But Eqs.(\ref{behera3}) are based on the assumption that all the dependence on $(x', y', z')$ appears \textit{explicitly} in Eq.(\ref{behera2}). This is not the case, because although $k , l , m$ are constants and independent of the coordinates, the same need not be true of $c'$. In fact, the above quote from the paper of Behera and Mukhopadhyay directly contradicts Eqs.(\ref{behera3}). Indeed, admitting that $c'$ may vary in different directions \textit{means} that it contains some dependence on $(x', y', z')$. A proper substitution of the coordinate transformations Eq.(\ref{behera1}) into the wavefront equation Eq.(\ref{waveSp}) requires also substituting the adequate expressions into the function $c'$. This may turn the functional dependence of Eq.(\ref{behera2}) on $(x , y , z)$ into just about anything, however. Without any knowledge of how $c'$ depends on directions, i.e., on the variables $(x', y', z')$, we cannot proceed to Eqs.(\ref{behera3}). Without this, we cannot prove that $c' = c$, contrary to the claim of Behera and Mukhopadhyay.

Thus, the derivation of Eqs.(\ref{behera3}) is based on an implicit assumption, namely, that $c'$ is independent of the direction of propagation of the light signal. This assumption is \textit{a} second postulate, clearly related to Einstein's postulate, though logically weaker (the analysis of this postulate is part of an upcoming work \citep{drory2}).

\section{The derivation of Sfarti}
\label{sec:sfarti}

A different attempt to derive the constancy of the speed of light from the principle of relativity appears in a series of papers by A. Sfarti \citep{sfarti1, sfarti2, sfarti3}. This work has been cited by others who attempted to rederive SR from fewer postulates. Thus, Sela et al. have Sfarti presenting a kind of unorthodox relativity, although Sfarti actually attempts to derive the standard Lorentz transformation \citep[p. 508]{sela}, while Sfarti himself mentions that he discussed his derivation with N. David Mermin \citep{sfarti3}. In view of the relative attention Sfarti's work garnered, therefore, I believe it is useful to analyse his argument in detail and show exactly where it goes wrong. 

The set-up is identical in all three papers and follows Einstein's clock synchronization procedure \citep{einstein1}. Sfarti imagines that a light pulse is emitted from the origin A' of a reference frame S' at the time $t'_0 = 0$ and reflected at a later time $t'_B$ by a mirror placed at a point B', whose space coordinate is $x'_B$. The reflected signal returns to the origin at a time $t'_A$. The clocks at A' and B' are said to be synchronized if $t'_A - t'_B = t'_B - t'_0$. This procedure guarantees that when measured, the speed of the light signal going from A' to B', which I shall denote $c_0$, is the same as that of the signal reflected from B' to A', since speed measurements depend on the clock synchronization procedure. 

By the principle of relativity, this synchronization procedure must also be valid for any other inertial system S. Hence, another observer S, when synchronizing \textit{his} clocks, will find according to them that any light signal \textit{he} sends from \textit{his} origin A will travel to any point B in the same time the reflected signal travels back from B to A. This procedure does not in itself guarantee that the speed $c$ that S measures for a signal emitted by a source at rest in S' (which is therefore in motion with respect to S), is identical to the speed $c_0$ that S' measures. This is because the synchronization procedure only involves sources at rest in the frame of the observer.  Sfarti therefore endeavors to \textit{prove} that $c = c_0$, and thus to show that the second postulate can be derived from the first. 

Consider now the light signal that S' emits from his origin A' when synchronizing his clocks, from the perspective of the observer S. From the point of view of S this same signal reaches the point B' at a time $t_B$ and returns to the origin of S', point A', at a time $t_A$. Let us denote by $r_{AB}$ the distance that S measures between A and B (which is not necessarily the same distance that S' measures). Next, denote by $c_+$ the speed of the light signal sent from A' to B' as S measures it, and by $c_-$ the speed of the signal reflected from B' to A'. Then, because the points A' and B' move with respect to S with a speed $v$, the time $t_+$ it takes the signal to reach the point B' (as measured by S), is:
\begin{equation}
\label{sfar2}
c_+t_+ = r_{AB} + vt_+
\end{equation}
whereas the time $t_-$ it takes the reflected signal to reach the point A' (which has advanced in the meantime) is
\begin{equation}
\label{sfar3}
c_-t_- = r_{AB} - vt_-
\end{equation}

In both \citep{sfarti1} and \citep{sfarti2}, these correspond to Eqs.(2) and (3). Sfarti's equations differ from Eqs.(\ref{sfar2})-(\ref{sfar3}) in one crucial detail, however: he does not distinguish between the speeds of light going right and left, i.e., he implicitly assumes that $c_+ = c_-$. For signals sent by S from his own frame, e.g., from a light source attached to his own origin A, this is necessarily true because of the clock synchronization procedure. But Sfarti's relation is supposed to apply to the light signal emitted by S', i.e., from a source that is attached to the origin A', and that therefore moves at speed $v$ with respect to the S frame. Nothing in the clock synchronization procedure applies to this case. 

Indeed, the clock synchronization procedure that Sfarti adopts, following Einstein, would be perfectly acceptable to a classical physicist, who would nevertheless certainly not consider the claim  $c_+ = c_-$ to be obvious. In a (Newtonian-like) emission theory, a light particle colliding elastically with the mirror at point B' would indeed be reflected with a speed identical to that it had before reaching B' in the mirror's rest frame (S'). Yet for S, that light signal would be emitted from a source in motion, and its speed would then be increased by the speed of the source. On the other hand, the speed of the reflected light signal, being bounced off a mirror that is itself moving at speed v, would be lower. Thus in this case, we would expect $c_+ > c_-$. 

Similarly, in an ether wave theory of light the speed of the signal would depend on the velocity of the frame S' with respect to the ether. We might then obtain either $c_+ > c_-$, $c_+ < c_-$, or $c_+ = c_-$, depending on whether S' moves, with respect to the ether, with positive, negative or zero velocity, respectively. 

The point is that all these options (and others, of course) are still compatible with the clock synchronization procedure. Thus, for Sfarti to take $c_+ = c_-$ represents an additional and unfounded assumption, which is clearly similar to the assumption of the constancy of the speed of light, i.e., the very principle that Sfarti claims he can dispense with. 

Although there is no explicit mention of this in his papers, it seems that Sfarti recognized this fact later. Consequently, he tried to \textit{prove} with this same set-up that $c_+ = c_-$ \citep{sfarti3}. The supposed proof is erroneous, however, as shown in \citep{drorycan}.

In fact, however, Sfarti's assumption is stronger than just $c_+ = c_-$, because in all three papers considered here, he rotates the light source of S' by $90^0$ and consider a light signal propagating along the y' axis (this is essentially Einstein's light clock thought experiment). From the point of view of S, the light signal appears to run along some diagonal path. In analysing this phenomenon, Sfarti uses the speed $c = c_+ = c_-$ for the speed of this signal along the diagonal path in the S frame. This is a stronger assumption, similar to the one made by Behera and Mukhopadhyay. It assumes that  a light signal moving in any direction whatsoever in the frame S will do so at the same speed, even though it is emitted from a source moving in the x-direction. There is nothing to support this assumption, however [see \citep{drorycan} for more details].

Yet even if we were to grant Sfarti's contention that $c_+ = c_- = c$, the remainder of the argument, which purports to show that $c = c_0$, is incorrect. Sfarti's Eq.(26) in \citep{sfarti1} [or identically Eq.(10) in \citep{sfarti2}], reads, in the present notations,
\begin{eqnarray}
\label{sfar26}
x = \dfrac{1}{\sqrt{1 - \dfrac{v^2}{c^2}}}\left(x' + vt'\right) \nonumber \\
t = \dfrac{1}{\sqrt{1 - \dfrac{v^2}{c^2}}}\left(t' + \dfrac{v}{c^2}x'\right)
\end{eqnarray}
Recall that in this equation, $c$ represents the velocity of light that S measures when observing a signal emitted from the origin of the frame S'. According to Sfarti's (unprovable) assumption that $c_+ = c_-$ and its extension, this is the velocity of light emitted in any direction whatsoever from a source that moves with speed $v$ in relation to S. In the next step, Sfarti algebraically inverts the equations to obtain Eq.(27) in \citep{sfarti1} [or Eq.(11) in \citep{sfarti2}], rewritten here as:
\begin{eqnarray}
\label{sfar27}
x' = \dfrac{1}{\sqrt{1 - \dfrac{v^2}{c^2}}}\left(x - vt\right) \nonumber \\
t' = \dfrac{1}{\sqrt{1 - \dfrac{v^2}{c^2}}}\left(t - \dfrac{v}{c^2}x\right)
\end{eqnarray}
Now comes the crucial - and incorrect - step. Sfarti claims that by the principle of relativity one can reverse the roles of S and S', and obtain directly from Eq.(\ref{sfar26}) the relation
\begin{eqnarray}
\label{sfar28}
x' = \dfrac{1}{\sqrt{1 - \dfrac{v^2}{c_0^2}}}\left(x - vt\right) \nonumber \\
t' = \dfrac{1}{\sqrt{1 - \dfrac{v^2}{c_0^2}}}\left(t - \dfrac{v}{c_0^2}x\right)
\end{eqnarray}
If this were correct, comparison with Eq.(\ref{sfar27}) would yield immediately that $c = c_0$, which is what Sfarti seeks to prove. Unfortunately, Eq.(\ref{sfar28}) is based on a physical mistake.  

To see why, we must be very explicit about the meaning of $c_0$ and $c$. $c_0$ is measured in the rest frame of the source (S'), and  must therefore be a simple number. On the other hand, $c$ represents the speed of a light signal emitted by a source that moves with velocity $v$ with respect to the observer (S). Logically, there are two possibilities. Either $c$ depends on $v$ or it doesn't. 

Now if $c$ does not depend on $v$, it must also represent the velocity of light emitted by a source at rest in S. Here is where the principle of relativity comes into play. In the frame S', an emitter at rest produces a signal that propagates with speed $c_0$, whereas in the frame S, a signal emitted by an emitter \textit{at rest in S} propagates with a speed $c$. Since the principle of relativity requires both frames to be equivalent, we must have $c_0 = c$, which is precisely what Sfarti seeks to prove. Indeed, Sfarti's contention that the reversal of roles of S and S' produces Eq.(\ref{sfar28}) is a formal way of saying the very same thing. 

That is no proof, however, because the conclusion only holds if $c$ is independent of $v$. But if we assume this, the entire argument is redundant because having $c$ independent of $v$ is precisely the principle of constancy of the speed of light. The problem is that there is another possibility, namely, that $c$ is some function of $v$, the speed of the emitter.

Now if that is the case, we can rewrite $c = c(v)$ to make the relationship more explicit. In this case, we only have $c_0 = c(v = 0)$, in accordance with the principle of relativity, which requires that a signal emitted from a source at rest in S propagates with the same speed as a signal emitted from a source at rest in S'. Let us now rewrite Eqs.(\ref{sfar27}) accordingly as:

\begin{eqnarray}
\label{sfar27b}
x' = \dfrac{1}{\sqrt{1 - \dfrac{v^2}{c(v)^2}}}\left(x - vt\right) \nonumber \\
t' = \dfrac{1}{\sqrt{1 - \dfrac{v^2}{c(v)^2}}}\left(t - \dfrac{v}{c(v)^2}x\right)
\end{eqnarray}

Now consider Eqs.(\ref{sfar26}) and invert the roles of S and S'. With respect to S, S' moves at velocity $v$, but with respect to S', S moves at the velocity $-v$. Therefore the inversion leads not to Eqs.(\ref{sfar28}), as Sfarti claims, but rather to the following set of equations:

\begin{eqnarray}
\label{sfar28b}
x' = \dfrac{1}{\sqrt{1 - \dfrac{v^2}{c(-v)^2}}}\left(x - vt\right) \nonumber \\
t' = \dfrac{1}{\sqrt{1 - \dfrac{v^2}{c(-v)^2}}}\left(t - \dfrac{v}{c(-v)^2}x\right)
\end{eqnarray}

The principle of relativity will only yield the requirement that $c^2(v) = c^2(-v)$, but not the required $c(v) = c(0)$. Note that the relation $c^2(v) = c^2(-v)$ is a variant of the condition $c_+ = c_-$. In the first case, it is the emitter's motion that is rotated by $180^0$ with respect to the direction of propagation of the signal, while in the second it is the direction of propagation that is rotated with respect to the emitter's direction of motion. By the principle of relativity, these two processes must be equivalent, since only the relative rotation can matter. This means that the relation $c^2(v) = c^2(-v)$ is just a restatement of the assumption $c_+ = c_-$ and yields nothing new. Since the condition $c_+ = c_-$ represents an unproven assumption, and Sfarti's attempt to prove it, \citep{sfarti3}, is in fact incorrect, \citep{drorycan}, we have gained nothing. 

The failure here is exactly the same as in the work of Behera and Mukhopadhyay, namely, a hidden assumption that light propagates isotropically, independently of the velocity of its source.

\section{Conclusions}
\label{sec:conclusions}

The derivations of the generalized Lorentz transformations seem to imply that the role of the second postulate in the theory of special relativity is merely the report of the numerical value of a free parameter, $k$, which appears in the equations. 

I have argued, to the contrary, that the generalized Lorentz equations cannot be said to represent special relativity insofar as special relativity is a theory that opposes Newtonian mechanics in its world-view. It does so by denying the possibility that $k = 0$, which immediately implies that simultaneity is relative and that distances and durations are not frame-invariant. These facts in turn  influence the manner in which experiments are performed. The generalized Lorentz transformations offer no hint whether $k$ vanishes or not, however, and are not therefore in opposition to Newtonian mechanics. Instead, they include it as a possibility (not merely as a low-velocity approximation as special relativity does.) Thus, these transformations do not in fact entail a definite world-view regarding the nature of simultaneity, distances or duration.

Some derivations, by Sfarti and Behera and Mukhopadhyay, try to obtain not only the generalized Lorentz transformations but also the actual value of $k$. These derivations are erroneous, however, and make hidden use of the second postulate at key moments.

\begin{appendix}
\section{Appendix}

Let us assume the postulates E-G of section \ref{sec:galileo}, namely space-time homogeneity, the principle of relativity and the principle of the constancy of the speed of light, but not isotropy. 

As shown in section \ref{sec:trans}, space-time homogeneity and the principle of relativity require the transformations to be linear, so we have, as in Eq.(\ref{stage1}),

\begin{eqnarray}
\label{stageA1}
x' = \gamma\left( x - vt \right) \nonumber \\
t' = \alpha \left(t - \epsilon x \right),
\end{eqnarray}
where $\alpha, \epsilon , \gamma$ are as yet undetermined constants, which may depend on the parameter $v$. 

Instead of proceeding with isotropy, let us now impose only the condition that the velocity of light be identical in all frames. In the frame S, any light signal emitted from the origin and propagating along the x-axis will be described as
\begin{equation}
\label{eq:Asignal}
x = \pm ct
\end{equation}
where the $\pm$ sign, which corresponds to signals emitted in the positive or the negative x-direction, follows from the requirement that \textit{any} light signal must travel at the speed $c$, irrespective of its direction.

Inserting Eq.(\ref{eq:Asignal}) into the transformations, Eqs.(\ref{stageA1}), we obtain:
\begin{eqnarray}
\label{Atranslight}
x' = \gamma \left(\pm c - v\right) t \nonumber \\
t' = \alpha \left(1 \mp c \epsilon \right) t
\end{eqnarray}

The principle of the constancy of the speed of light requires that the same light signal be described, in the $S'$-frame, as $x' = \pm ct'$. Inserting into this relation the transformations Eqs.(\ref{Atranslight}), then separating the two cases (motion along the positive and the negative x-axis), we obtain:
\begin{eqnarray}
\label{Apm}
\gamma \left(c - v \right) = c \alpha \left( 1 - c \epsilon \right) \\
\label{prule}
\gamma \left(c + v \right) = c \alpha \left( 1 + c \epsilon \right) 
\end{eqnarray}
which yields immediately
\begin{eqnarray}
\alpha = \gamma  \nonumber \\
\epsilon = \dfrac{v}{c^2}
\end{eqnarray}

The coordinate transformations now become
\begin{eqnarray}
\label{eq:Atrans1}
x' = \gamma \left( x - vt \right) \nonumber \\
t' = \gamma \left( t - \dfrac{v}{c^2}x \right)
\end{eqnarray}

The form of the remaining parameter $\gamma$, which clearly may depend on $v$, is now fixed by the requirement that these transformations should form a group.

Introduce the more convenient rapidity parameter, $\varphi$, defined by the relation
\begin{equation}
\dfrac{v}{c} = tanh \, \varphi
\end{equation}

Furthermore, define a new function, $g(\varphi)$ so that:
\begin{equation}
\label{Adefg}
\gamma =  g(\varphi) cosh \, \varphi \,
\end{equation}

The transformations Eqs.(\ref{eq:Atrans1}) now become, in matrix form:
\begin{equation}
\label{Amatrix}
\left( \begin{array}{c}
x' \\
ct' \\ 
\end{array}\right)
= g(\varphi) \left(\begin{array}{cc}
 cosh \, \varphi & - sinh \, \varphi \\ 
-sinh \, \varphi & cosh \, \varphi 
\end{array} \right)
\left( \begin{array}{c}
x \\
ct \\
\end{array} \right)
\end{equation}

Let us assume then that a frame S' moves with respect to S with velocity $v$ and rapidity $\varphi$, and that a third frame, S'', moves with velocity $u$ and rapidity $\psi$ with respect to the S'-frame. The transformation from S' to S'' now reads
\begin{equation}
\label{eq:Atrans2}
\left( \begin{array}{c}
x'' \\
ct'' \\ 
\end{array}\right)
= g(\psi) \left(\begin{array}{cc}
 cosh\, \psi & - sinh\, \psi \\ 
-sinh \, \psi & cosh\, \psi 
\end{array} \right)
\left( \begin{array}{c}
x' \\
ct' \\
\end{array} \right)
\end{equation}

Combining these transformations with the transformations from S to S', Eq.(\ref{Amatrix}), we obtain the relations
\begin{equation}
\label{eq:Acombo1}
\left( \begin{array}{c}
x'' \\
ct'' \\ 
\end{array}\right)
= g(\psi)\, g(\varphi) \left(\begin{array}{cc}
 cosh \, \psi & - sinh \, \psi \\ 
-sinh \, \psi & cosh \, \psi 
\end{array} \right)
\left(\begin{array}{cc}
 cosh \, \varphi & - sinh \, \varphi \\ 
-sinh \, \varphi & cosh \, \varphi 
\end{array} \right)
\left( \begin{array}{c}
x \\
ct \\
\end{array} \right)
\end{equation}
or, using basic identities from hyperbolic geometry, 
\begin{equation}
\label{eq:Acombo}
\left( \begin{array}{c}
x'' \\
ct'' \\ 
\end{array}\right)
= g(\varphi)\, g(\psi) \left(\begin{array}{cc}
 cosh \left(\varphi + \psi \right) & - sinh \left(\varphi + \psi \right) \\ 
-sinh \left(\varphi + \psi \right) & cosh \left(\varphi + \psi \right)
\end{array} \right)
\left( \begin{array}{c}
x \\
ct \\
\end{array} \right)
\end{equation}

The group property requires that this should be identical with the direct transformation from S to S'',
\begin{equation}
\label{eq:Adirect}
\left( \begin{array}{c}
x'' \\
ct'' \\ 
\end{array}\right)
= g(\xi) \left(\begin{array}{cc}
 cosh\, \xi & - sinh \, \xi \\ 
-sinh\, \xi & cosh\, \xi
\end{array} \right)
\left( \begin{array}{c}
x \\
ct \\
\end{array} \right)
\end{equation}
where $\xi$ is the rapidity of the S''-frame with respect to the S frame. 

Comparison of the matrix on the right hand side with the matrix in the combined transformation, Eq.(\ref{eq:Acombo}), shows immediately that we must have
\begin{equation}
\label{eq:Aveloadd}
\xi = \varphi + \psi  \quad \Longrightarrow \quad  w = \dfrac{v + u}{1 + \dfrac{v \, u}{c^2}}
\end{equation}
where $w$ is the velocity of the S''-frame with respect to the S frame, $w = tanh \, \xi$. Eq.(\ref{eq:Aveloadd}) is the standard relativistic velocity addition formula. Thus, this formula does \textit{not} require isotropy. It follows from only the principle of relativity and the principle of the constancy of the speed of light, along with space-time homogeneity.

Comparison of the multiplicative factor in front of the matrix on the right hand sides of Eqs.(\ref{eq:Acombo}) and (\ref{eq:Adirect}) yields the relation
\begin{equation}
\label{eq:Aq}
g(\xi) \equiv g(\varphi + \psi) = g(\varphi) \, g(\psi)
\end{equation}
Note that if we choose $\psi = 0$ (or equivalently $u = 0$) in this equation, we find that $g(0) = 1$. The Lorentz transformations correspond to $g(\varphi) = 1$ identically, but the general solution to Eq.(\ref{eq:Aq}) is $g(\varphi) = e^{2\eta \varphi}$, where $\eta$ is a constant.

The proof is straightforward. Subtracting $g(\varphi)$ from both sides of Eq.(\ref{eq:Aq}) and then dividing by $\psi$, we can rewrite the equation as
\begin{equation}
\dfrac{g(\varphi + \psi) - g(\varphi)}{\psi} = \dfrac{g(\varphi)\left[g(\psi) - 1\right]}{\psi} = g(\varphi)\left[\dfrac{g(\psi) - g(0)}{\psi} \right]
\end{equation}

Now let $\psi \longrightarrow 0$. We immediately have that:
\begin{equation}
\label{eq:Aderivg1}
\dfrac{d g}{d\varphi}(\varphi) = g(\varphi) \, \dfrac{d g}{d \varphi}(0)
\end{equation}
The derivative of $g(\varphi)$ at $\varphi = 0$ is a constant that we can denote as $2 \eta$. Thus, we obtain the equation
\begin{equation}
\label{eq:Aderivg}
\dfrac{d g}{d \varphi} (\varphi) = 2 \eta \,g(\varphi)
\end{equation}
which has the immediate solution
\begin{equation}
g(\varphi) = A \, e^{2 \eta \, \varphi}
\end{equation}
where A is a constant. Since $g(0) = 1$, however, we must have $A = 1$ and hence $g(\varphi) = e^{2 \eta \, \varphi}$.

Returning now to the definition of $g(\varphi)$, Eq.(\ref{Adefg}), we have finally
\begin{equation}
\label{Aq0}
\gamma = e^{2\eta \, \varphi} cosh \, \varphi = \left( \dfrac{c + v}{c - v}\right)^{\eta} \dfrac{1}{\sqrt{1 - \dfrac{v^2}{c^2}}}
\end{equation}
Here we made use of the identity
\begin{equation}
\dfrac{v}{c} = tanh \, \varphi = \dfrac{e^{2\varphi} - 1}{e^{2\varphi} + 1} \qquad \Longrightarrow  \qquad e^{2 \varphi} = \dfrac{c + v}{c - v} 
\end{equation}
It is important to note that $\eta$ does not depend on $v$ and is therefore frame-invariant.

 Finally, consider the transverse axes. Let a light signal be emitted in a diagonal direction in the xy plane from the origin of the frame S at a time $t_0 = 0$. The position of a point on the wavefront is $(ct, x, y, 0)$ when observed in the S frame, and $(ct', x', y', 0)$ when observed in the S' frame. Since the speed of light is invariant, we must have that 

\begin{eqnarray}
\label{Awavefront}
x^2 + y^2 = c^2t^2 \nonumber \\
x'^2 + y'^2 = c^2t'^2
\end{eqnarray}

Substituting the transformations for $x'$ and $t'$ from Eq.(\ref{eq:Atrans1}) and comparing with the first of Eqs.(\ref{Awavefront}), we obtain

\begin{equation}
y'^2 = c^2t'^2 - x'^2 =  \gamma  \left(1 - \dfrac{v^2}{c^2}\right) y^2
\end{equation}

Using Eq.(\ref{Aq0}) for $\gamma$, we obtain the final form of the transformations:
\begin{eqnarray}
\label{Afinalx}
x' = \left( \dfrac{1 + \beta}{1 - \beta} \right)^{\eta} \dfrac{x - \beta c t}{\sqrt{1 - \beta^2}}  \\
\label{Afinaly}
y' = \left( \dfrac{1 + \beta}{1 - \beta} \right)^{\eta} y \\
\label{Afinalz}
z' = \left( \dfrac{1 + \beta}{1 - \beta} \right)^{\eta} z \\
\label{Afinalt}
ct' = \left( \dfrac{1 + \beta}{1 - \beta} \right)^{\eta} \dfrac{ct - \beta x}{\sqrt{1 - \beta^2}}
\end{eqnarray}
where $\beta = \dfrac{v}{c}$, as usual. The transformation for $z'$ is obtained in the same way as the transformation for $y'$.

The above derivation shows that $\eta$ is independent of position, time and frame velocity, but we shall now see that its value depends in general on the choice of the x-axis, or equivalently, on the orientation of the frame. This is why this exponent expresses the degree of spatial isotropy.

Let S' be a frame moving at a velocity $v$ in the positive x direction with respect to the frame S. The x-axis and x'-axis are parallel. Consider an event E happening at a space-time point $\left( ct, x \right)$ in the S-frame. In the S'-frame, the position of the event is given by Eq.(\ref{Afinalx}).

Now let us rotate the frames by $\pi$ radians. Denote the rotated S-frame by $S_R$ and the rotated S'-frame by $S'_R$. We do not reverse the direction of motion of the frame S', and consequently,  $\beta_R = - \beta$. The event E now takes place on the negative x-axis, so that $x_R = -x$ and $x'_R = - x'$. Clearly we have that 

\begin{equation}
\label{AxR1}
x'_R = \left( \dfrac{1 + \beta_R}{1 - \beta_R}\right)^{\eta_R} \dfrac{x_R - \beta_R c t}{\sqrt{1 - \beta_R^2}}
\end{equation}

Substituting the values of $x_R$ and $\beta_R$ into Eq.(\ref{AxR1}) yields now

\begin{equation}
\label{AxR2}
-x'= x'_R = \left( \dfrac{1 - \beta}{1 + \beta}\right)^{\eta_R} \dfrac{-x + \beta c t}{\sqrt{1 - \beta^2}}
\end{equation}

Comparison with Eq.(\ref{Afinalx}) shows immediately that 

\begin{equation}
\label{Akr}
\eta_R = - \eta
\end{equation}

Thus, we see that if the exponent $\eta$ does not vanish, its value depends on the orientation of the $x$-axis, which means that not all directions are equivalent. 

Since $\eta$ changes sign under a rotation of $\pi$, there must be at least one specific direction in space for which $\eta = 0$. Along this axis, standard special relativity must be valid without any alterations. Let us then denote by $\theta$ the angle of any other direction with respect to this preferred axis. Eq.(\ref{Akr}) can now be rewritten as

\begin{equation}
\label{Akappatheta}
\eta(\theta + \pi) = - \eta (\theta)
\end{equation}
which is necessarily a symmetry of the anisotropy exponent.

\end{appendix}

\end{document}